\documentclass[twocolumn, secnumarabic,amssymb, amsmath,nobibnotes, aps, pre]{revtex4}
\usepackage{graphicx}
\usepackage{color}
\usepackage[ansinew]{inputenc}
\usepackage{amsmath}
\usepackage{amsfonts}
\usepackage{amssymb}
\usepackage{dcolumn}
\usepackage{float}

\begin{document}

\title{Random Sequential Adsorption of Oriented Superdisks}

  \author{Oleksandr Gromenko}
  \affiliation{Department of Physics, Clarkson University, Potsdam, NY 13699, USA}
  \author{Vladimir Privman}
  \affiliation{Department of Physics, Clarkson University, Potsdam, NY 13699, USA}


 \begin{abstract}
In this work we extend recent study of the properties of the
dense packing of ``superdisks,'' by Y.\ Jiao,\ F.\ H.\ Stillinger\ and
S.\ Torquato,\ Phys.\ Rev.\
Lett.\ \textbf{100},\ 245504\ (2008), to the
jammed state formed by these objects in random sequential adsorption.
The superdisks are
two-dimensional shapes bound by the curves of the form
$|x|^{2p}+|y|^{2p}=1$, with $p >0$. We use Monte Carlo
simulations and theoretical arguments to establish that $p=1/2$ is
a special point at which the jamming density, $\rho_J(p)$, has a
discontinuous derivative as a function of $p$. The existence of
this point can be also argued for by geometrical arguments.
 \end{abstract}
\maketitle

There has been recent interest \cite{Torquato}, \cite{Shelke}, \cite{Mizoshita},
\cite{Schock}, \cite{Saettel} in the problem of geometrical packing and
surface deposition of noncircular objects in two
dimensions (2D). This problem is intriguing from the theoretical
point of view. In addition, it finds applications in studies of
design and control of prepatterned surfaces with special
properties. New capabilities to pattern surfaces at the
nanoscale, and use particles of nanosizes, have promise for
development of novel biosensors and detectors, applications
in electronics \cite{Palma}, \cite{Berresheim}, catalysis
\cite{Thomas}, and optics \cite{Sacana},
\cite{Yu}.

\begin{figure}[t]
\begin{center}
\includegraphics[width=3.0 true in]{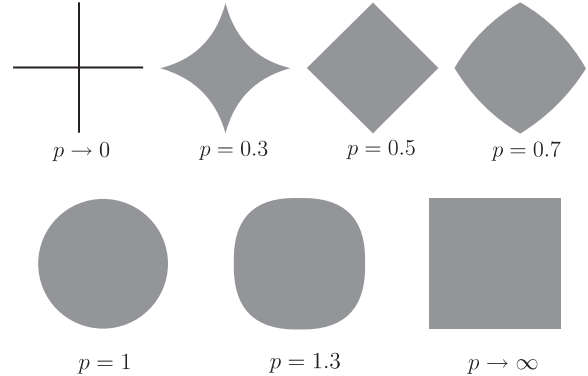}
\caption{Superdisks shapes for different values of the
deformation parameter $p$.}
\label{Fig1}
\end{center}
\end{figure}

Recently, an interesting study was reported \cite{Torquato} of the
densest possible packing of (oriented) ``superdisks'' defined by
$|x|^{2p}+|y|^{2p} \leq 1$. These shapes are illustrated in Fig.\
\ref{Fig1}. In particular, numerical evidence for $0 \leq p \leq
1$ (where the $p = 0$ shapes are defined as a limit which yields
crosses) suggests \cite{Torquato} that the point $p=1/2$ separates
different closed-packed structures. Note that $p=1/2$ also
separates the convex and concave shapes, as shown in Fig.\ \ref{Fig1}.

Particle deposition at surfaces is typically irreversible, and for
a theoretical description of their adsorption one can use the
random sequential adsorption (RSA) model. The RSA model, as well
as its various modifications, finds applications and has been
extensively studied \cite{Evans},
\cite{Privman1}, \cite{Ramsden}, \cite{Privman33}, \cite{Privman55},
\cite{Tassel}, \cite{Torquato1}, \cite{Privman2}, \cite{Wang}, \cite{Cadilhe2}, 
\cite{Cadilhe}, \cite{Nielaba}, \cite{Wang2}, \cite{Privman8},
\cite{Privman5}, \cite{Gromenko}, \cite{Gromenko2}, \cite{Gonzalez}.

However, most RSA studies have been carried out for spherical
(circular) and other simply shaped objects. The approach to the jamming
density, $\rho_J$, in RSA processes is described by the standard
Pomeau \cite{Pomeau} and Swendsen \cite{Swendsen} conjuncture
which gives the asymptotic results for oriented squares and for disks,
which are in agreement with Monte Carlo (MC) simulation results
\cite{Brosilow}, \cite{Privman3} (oriented squares), and
\cite{Torquato1}, \cite{Hinrichsen} (disks). However, 
for non-oriented squares evidence has been
reported \cite{Viot} that this conjuncture might not work.
Moreover, the asymptotic behavior of the deposit density for
objects with concave shapes on continuum substrates has not been
studied. Studies of RSA of objects with zero area, such as rods,
circular arcs, etc., have reported interesting features of the
jamming coverage \cite{Khandkar1}, \cite{Khandkar2}.

\begin{figure}[b]
\begin{center}
\includegraphics[width=3.0 true in]{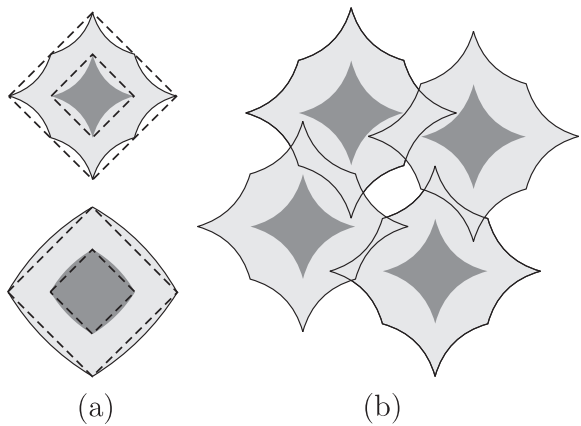}
\caption{(a) Superdisks (dark shapes) with their
exclusion areas (lighter shapes). Upper panel: a concave 
superdisk for $p=0.3$. Lower panel: a convex
superdisk for $p=0.7$. The dashed lines mark the $p=0.5$ squares and their
exclusion areas. (b) A typical configuration of concave
superdisks near the jammed state, with at most a single
additional superdisk deposition possible with its center 
landing in the central unshaded area.}
\label{Fig2}
\end{center}
\end{figure}

In this work we consider RSA of oriented
superdisks in two dimensions. We use a grid-type MC
algorithm \cite{Brosilow} which is particularly suitable for
evaluating the density of the jammed state, because it efficiently treats
deposition in small remaining vacancy areas close
to jamming; see Fig.\ \ref{Fig2}. Similar to the dense-packing
results \cite{Torquato}, we find that $p=1/2$ is also a special point
for the jammed state of RSA. In addition to numerical evidence,
this conclusion will also be substantiated by geometrical arguments.

A superdisk is a 2D case of the surface of a
$d$-dimensional superball \cite{Torquato}. A superball is defined as the
volume of the Euclidean space bounded by the surface
$|x_1|^{2p}+|x_2|^{2p}+...+|x_d|^{2p}=1$, where $x_i$ are the
Cartesian coordinates and $p$ is the deformation parameter.
Supeballs have full rotational symmetry only when $p=1$
(when they became hyperspheres).

For $0 < p < 1/2$ superdisks are
concave and for $1/2 < p < \infty$ they are convex. The
$p \to 0$ superdisk is reduced to cross, the $p=1/2$ and
$p=\infty$ shapes are squares, and the $p=1$ shape is a circle.

The reason that we focus on the point $p=1/2$ is that, with the advent
of nanotechnology, and with proliferation of experiments on deposition of
proteins \cite{Ramsden}, \cite{Ramsden2}, \cite{Malmsten}, \cite{Gray}, \cite{Hyun} we expect 
that situations will be realized when the particle
shapes on the surface, change between concave and convex depending on
the physical and
chemical conditions of the environment. This might affect the
asymptotic approach to the jamming coverage (an issue
that requires a separate detailed study). As demonstrated here,
the change in the concavity also results in a nonanalytic 
behavior of the jamming coverage,
$\theta_J(p)$, at $p=1/2$.

In RSA, a superdisk can be deposited at a surface if it does
not overlap previously deposited particles. Such adsorption is a
nonequilibrium process, and therefore the deposited particle density
does not reach the maximal dense packing. Instead, it approaches
the jamming density, $\rho_J(p)$, at large times. This quantity,
the density of the deposited particles per unit area, is related
to the jamming coverage $\theta_J(p)\,$---$\,$the fraction of the covered
area$\,$---$\,$via $\theta_J(p)=A(p)\rho_J(p)$, where $A(p)$
is the superdisk area. The latter quantity is given by
\begin{equation}
\label{new}
A(p)=\frac{1}{p}\Gamma^2 \! \left(\frac{1}{2p}\right) \! \bigg / \! \Gamma \! \left(\frac{1}{p}\right)
\end{equation}
where $\Gamma(x)$ is the standard gamma function.
Since this function is analytic near $p=1/2$, the behavior of
$\theta_J(p)$ and $\rho_J(p)$ at $p=1/2$ is easily related. We
focus on $\rho_J(p)$, because it simplifies some notation below.

\begin{figure}[b]
\begin{center}
\includegraphics[width=3.0 true in]{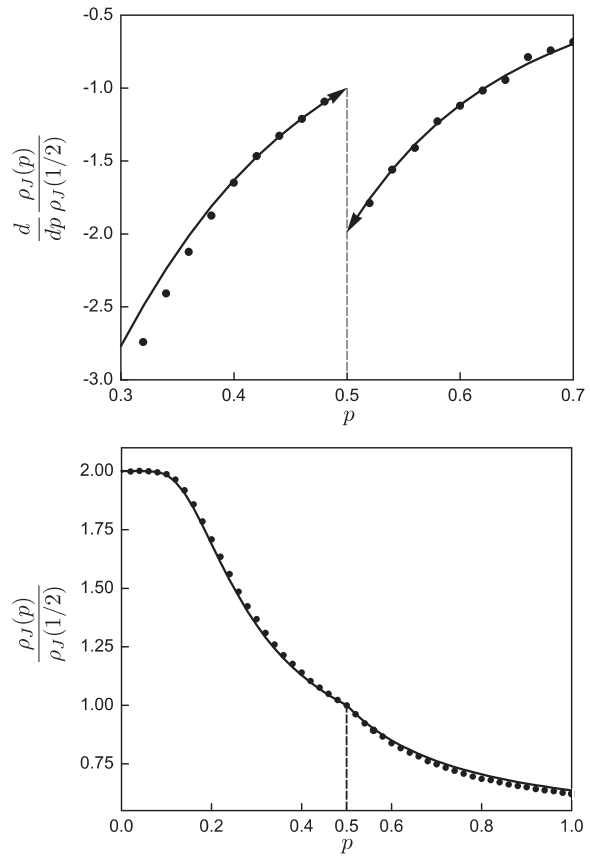}
\caption{Lower panel: normalized jamming density of the
superdisks, $\rho_J(p)/\rho_J(1/2)$, as a function of the deformation
parameter $p$. Upper panel: the $p$-derivative of the 
normalized jamming density near the special
point $p=1/2$. The symbols are the results of our MC simulations,
whereas the solid lines show the approximation (\ref{theory}).}
\label{Fig3}
\end{center}
\end{figure}

In our MC simulations we used an algorithm originally 
introduced in \cite{Brosilow},
which allows to simulate the formation of the jamming state, and
to estimate $\rho_J$, using minimal computer resources. We used a
square system of size $500D \times 500D$ with periodic boundary
conditions, where $D$ is the ``diameter'' of the superdisks
along the $x$ and $y$ axes, equal 2 in our dimensionless units. 
Each value of $\rho(p)$ was
obtained by averaging over 1000 independent runs. The
maximum fractional uncertainty in our simulation was estimated as
$\Delta \rho_J(p)/\rho_J(p)\simeq 0.00223$. Specifically, for the squares
$(p=1/2)$ and disks $(p=1)$ we obtained the estimates
$\theta_J(p=1/2)=0.5620\pm 0.0001$ and $\theta_J(p=1)=0.5468\pm
0.0005$, which are consistent with the values reported in
\cite{Brosilow}, \cite{Privman3}, \cite{Torquato1}, \cite{Hinrichsen}. The behavior of
the jamming density as a function of the deformation parameter is
shown in Fig.\ \ref{Fig3}.

Our data clearly indicate existence of a special point at $p=1/2$.
The $p$-derivative of the jamming density $\rho_J(p)$ at this point has a
discontinuity, similar to that mentioned in \cite{Torquato} for the
dense-packing density. In
order to understand the origin of this behavior in RSA, let us
consider the exclusion area of the superdisks, $S(p)$, defined as
the area within which it is impossible to
deposit another superdisk's center without overlap, Fig.\ \ref{Fig2}.
Unlike $A(p)$, the area $S(p)$ markedly changes its $p$-dependence for $p$ 
above and below the square-shape value of $1/2$. It is a continuous function of $p$,
but has a discontinuous derivative at $p=1/2$ (we give the expressions shortly).
Therefore, on dimensional grounds it is tempting to conjecture that the following 
relation provides a good qualitative {\it approximation\/} for the superdisk jamming density ratio,
\begin{equation}
\label{theory}
\frac{\rho_J(p)}{\rho_J(1/2)} \simeq \frac{S(1/2)}{S(p)}\,
\end{equation}
at least near $p=1/2$.

For superdisks, $s(p) \equiv S(1/2)/S(p)$ is given by following
relations,
\begin{equation}
\label{area2}
s_{+}(p)=
\frac{A(1/2)}{A(p)} = 2p \, \Gamma \! \left(\frac{1}{p}\right)  \Gamma^{-2} \! \left(\frac{1}{2p}\right)  \quad
 {\rm for} \quad p \geq \frac{1}{2}
\end{equation}
\begin{equation}
\label{area1} s_{-}(p)=\frac{2A(1/2)}{A(1/2)+A(p)} = \frac{2s_+(p)}{1+s_+(p)} \quad 
{\rm for} \quad 0< p \leq \frac{1}{2}
\end{equation}
Near $p=1/2$, the approximation (\ref{theory}) is a continuous
function of $p$, but has a jump in the $p$-derivative. In fact,
the jump in the derivative of $\rho_J(p)/\rho_J(1/2)$, given by our
exclusion area approximation, is in a reasonable agreement with the result
obtained by MC simulations presented in Fig.\ \ref{Fig3}.
The numerical values of the right and left $p$-derivatives of
$\rho_J(p)/\rho_J(1/2)$ at $p=1/2$ can be approximated by
\begin{equation}
\lim_{p= 1/2} \frac{ds_{+}(p)}{dp}=-2  \qquad
\lim_{p= 1/2} \frac{ds_{-}(p)}{dp}=-1
\end{equation}

In summary, in this work we demonstrated by numerical MC simulations, as well as by
approximate exclusion-area arguments, that the point at which the shape of the superdisks changes from concave to convex, is special not only in the geometric closed-packing properties, but also in the jammed-state properties in RSA. Future work will be focused on the dynamical simulations, to explore the approach to the jammed state, as well as on studies of unoriented superdisks.

The authors thank Prof.\ S.\ Torquato for instructive correspondence,
and acknowledge support of this research by the NSF under grant DMR-0509104.


\begin{thebibliography}{00}

\bibitem{Torquato}
Y.\ Jiao,\ F.\ H.\ Stillinger\ and\ S.\ Torquato,\ Phys.\ Rev.\
Lett.\ \textbf{100},\ 245504\ (2008).

\bibitem{Shelke}
P.\ B.\ Shelke,\ S.\ B.\ Ogale,\ M.\ D.\ Khandkar\ and\ A.\ V.\
Limaye,\ Phys.\ Rev.\ E\ \textbf{77},\ 066111\ (2008).

\bibitem{Mizoshita}
N.\ Mizoshita\ and\ T.\ Seki,\ Soft Matter \textbf{2},\ 157\
(2006).

\bibitem{Schock}
M.\ Sch\"{o}ck,\ R.\ Otero,\ S.\ Stojkovic,\ F.\ H\"{u}mmelink,
A.\ Gourdon,\ E.\ Lgsgaard, I.\ Stensgaard,\ C.\ Joachim\ and\ F.\
Besenbacher,\ J.\ Phys.\ Chem.\ B \textbf{110}, 12835 (2006).

\bibitem{Saettel}
N.\ Saettel,\ N.\ Katsonis,\ A.\ Marchenko,\ M.-P.\ Teulade-Fichou\
and\ D. Fichou,\ J.\ Mater.\ Chem.\ \textbf{15},\ 3175\ (2005).

\bibitem{Palma}
M.\ Palma,\ J.\ Levin,\ V.\ Lemaur,\ A.\ Liscio,\ V.\ Palermo,\
J.\ Cornil,\ Y.\ Geerts,\ M.\ Lehmann\ and P.\ Samor\`{i},\ Adv.\
Mater.\ \textbf{18},\ 3313\ (2006).

\bibitem{Berresheim}
A.\ J.\ Berresheim,\ M.\ M\"{u}ller\ and\ K.\ M\"{u}llen,\ Chem.\
Rev.\ \textbf{99},\ 1747\ (1999).

\bibitem{Thomas}
J.\ M.\ Thomas\ and\ W.\ J.\ Thomas,\ in\ {\it Principles\ and\
Practice\ of\ Heterogeneous\ Catalysis},\ VCH-Wiley: New York,
(1997).

\bibitem{Sacana}
S.\ Sacanna,\ L. Rossi,\ B.\ W.\ M.\ Kuipers\ and\ A.\ P.\
Philipse,\ Langmuir\ \textbf{22},\ 1822\ (2006).

\bibitem{Yu}
Y.\ Lu,\ Y.\ Yin\ and\ Y.\ Xia, Adv.\ Mater.\ \textbf{13},\ 415\
(2001).

\bibitem{Evans}
Review:\ J.\ W.\ Evans,\ Rev.\ Mod.\ Phys.\ \textbf{65},\ 1281\
(1993).

\bibitem{Privman1}
Review:\ M.\ C.\ Bartelt\ and\ V.\ Privman,\ Int.\ J.\ Mod.\
Phys.\ B\ \textbf{5},\ 2883\ (1991).

\bibitem{Ramsden}
Review:\ J.\ J.\ Ramsden,\ Chem.\ Soc.\ Rev.\ \textbf{24},\ 73\
(1995).

\bibitem{Ramsden2}
J.\ J.\ Ramsden,\ G.\ I.\ Bachmanova\ and A.\ I.\ Archakov,\ 
Phys.\ Rev.\ E \textbf{50},\ 5072\ (1994).

\bibitem{Malmsten}
M.\ Malmsten,\ ed.\ \textit{Biopolymers\ at\ Interfaces},\ (New\ York:\ Marcel\ Dekker,\ 2003).

\bibitem{Gray}
J.\ J.\ Gray, Curr.\ Opin.\ Struct.\ Biol.\ \textbf{14}, 110\ (2004).

\bibitem{Hyun}
J.\ Hyun,\ Y.\ Zhu,\ A.\ Liebmann-Vinson,\ T.\ P.\ Beebe\ and\ A.\ Chilkoti,\
Langmuir\ \textbf{17}, 6358 (2001).

\bibitem{Privman33}
Collection of review articles:\ {\it{}Adhesion\ of\ Submicron\ Particles\ on\ Solid\ Surfaces\/}, V.\ Privman,\ ed.,\ special\ volume\ of\ Colloids\
and\ Surfaces\ A\ \textbf{165},\ Issues\ 1-3,\ Pages\ 1-428\ (2000).

\bibitem{Privman55}
Collection of review articles:\ {\it{}Nonequilibrium\ Statistical\ Mechanics\ in\ One\ Dimension\/}, V.\ Privman,\ ed.\ (Cambridge University Press, Cambridge, 1997).

\bibitem{Tassel}
P.\ R.\ Van Tassel,\ P.\ Viot\ and\ G.\ Tarjus,\ J.\ Chem.\ Phys.\
\textbf{106},\ 761\ (1997).

\bibitem{Torquato1}
S.\ Torquato,\ O.\ U.\ Uche\ and\ F.\ H.\ Stillinger,\ Phys.\ Rev.\ E\ \textbf{74}, 061308 (2006).

\bibitem{Privman2}
Review:\ V.\ Privman,\ J.\ Adhesion\ \textbf{74},\ 421\ (2000).

\bibitem{Wang}
J.-S.\ Wang,\ P.\ Nielaba\ and\ V.\ Privman,\ Physica\ A\ \textbf{199},\ 527-538\
 (1993).

\bibitem{Cadilhe2}
N.\ A.\ M.\ Ara\'{u}jo,\ A.\ Cadilhe\ and\ V.\ Privman,\ Phys.\ Rev.\ E\ \textbf{77}, 031603 (2008).

\bibitem{Cadilhe}
N.\ A.\ M.\ Ara\'{u}jo\ and\ A.\ Cadilhe,\ Phys.\ Rev.\ E\ \textbf{73}, 051602\ (2006).

\bibitem{Nielaba}
P.\ Nielaba\ and\ V.\ Privman,\ Modern\ Phys.\ Lett.\ B
\textbf{6}, 533 (1992).

\bibitem{Wang2}
J.-S.\ Wang,\ P.\ Nielaba\ and\ V.\ Privman,\ Modern\ Phys.\
Lett.\ \textbf{B}\ \textbf{7}, 189 (1993).

\bibitem{Privman8}
M.\ C.\ Bartelt and V.\ Privman, Phys.\ Rev.\ A \textbf{44}, R2227 (1991).

\bibitem{Privman5}
A.\ Cadilhe, N.\ A.\ M.\ Ara\'{u}jo and V.\ Privman, J.\ Phys.\ Cond.\ Matter \textbf{19}, 065124 (2007).

\bibitem{Gromenko}
O.\ Gromenko,\ V.\ Privman\ and\ M.\ L.\ Glasser,\ J.\ Comput.\
Theor.\ Nanosci.\ \textbf{5},\ 2119\ (2008).

\bibitem{Gromenko2}
O.\ Gromenko\ and\ V.\ Privman,\ Phys.\ Rev.\ E\ \textbf{79},\ 011104\ (2009).

\bibitem{Gonzalez}
J.\ J.\ Gonzalez,\ P.\ C.\ Hemmer\ and\ J.\ S.\ H{\o}ye,\ Chem.\ Phys.\ \textbf{3},\ 288\ (1974).

\bibitem{Pomeau}
Y.\ Pomeau, J.\ Phys.\ A \textbf{13},\ L193 (1980).

\bibitem{Swendsen}
R.\ H.\ Swendsen,\ Phys.\ Rev.\ A \textbf{24},\ 504\ (1981).

\bibitem{Brosilow}
B.\ J.\ Brosilow,\ R.\ M.\ Ziff\ and\ R.\ D.\ Vigil,\ Phys.\
Rev.\ A\ \textbf{43},\ 631\ (1991).

\bibitem{Privman3}
V.\ Privman,\ J.-S.\ Wang\ and\ P.\ Nielaba,\ Phys.\ Rev.\ B\
\textbf{43},\ 3366\ (1991).

\bibitem{Hinrichsen}
E.\ L.\ Hinrichsen,\ J.\ Feder\ and\ T.\ J\o ssang\, J.\ Stat.\
Phys.\ \textbf{44},\ 793\ (1986).

\bibitem{Viot}
G.\ Tarjus and P.\ Viot,\ Phys.\ Rev.\ Lett.\ \textbf{67},\ 1875\
(1991).

\bibitem{Khandkar1}
P.\ B.\ Shelke,\ M.\ D.\ Khandkar,\ A.\ G.\ Banpurkar,\ S.\ B.\
Ogale\ and\ A.\ V.\ Limaye,\ Phys.\ Rev.\ E\ \textbf{75},
060601(R), (2007).

\bibitem{Khandkar2}
M.\ D.\ Khandkar,\ A.\ V.\ Limaye\ and S.\ B.\ Ogale,\ Phys.\
Rev.\ Lett.\ \textbf{84},\ 570\ (2000).
\end{thebibliography}
\end{document}